\documentclass[12pt]{article}
\usepackage{latexsym,amsfonts,amsmath,amsthm,amssymb}
\usepackage{pstricks}

\begin{document}

\title{Quantum-like model of cognitive decision making and information processing}
\author{Andrei Khrennikov\\
International Center for
Mathematical Modeling \\ in Physics and Cognitive Sciences\\
University of V\"axj\"o, S-35195, Sweden}

\maketitle

\abstract{ In this paper we present quantum-like (QL)
representation of the Shafir-Tversky statistical effect which is
well known in cognitive psychology. We apply so called contextual
approach. The Shafir-Tversky effect is considered as a consequence
of combination of a number of incompatible contexts which are
involved e.g. in Prisoner's Dilemma or in more general games
inducing the disjunction effect. As a consequence, the law of
total probability is violated for experimental data obtained by
Shafir and Tversky (1992) as well as Tversky and Shafir (1992).
Moreover, we can find a numerical measure of contextual
incompatibility (so called coefficient of interference) as well as
represent contexts which are involved in Prisoner's Dilemma (PD)
by probability amplitudes -- normalized vectors (``mental wave
functions''). We remark that statistical data from Shafir and
Tversky (1992) and Tversky and Shafir (1992) experiments differ
crucially from the point of view of mental interference. The
second one exhibits the conventional trigonometric ($\cos$-type)
interference, but the first one exhibits so called hyperbolic
($\cosh$-type) interference. We discuss QL processing of
information by cognitive systems, especially, QL decision making
as well as classical and QL rationality and ethics.}


\section{Introduction}

The author really wish that this paper would be readable by
psychologists, researchers working in cognitive science,
sociology, economics. Therefore an extended introduction contains
all basic ideas and methods of this paper. The corresponding
(simplified) mathematical considerations are placed at the end of
this paper, see section 7. On the other hand, one could not
totally escape the use of mathematics, since problems under
considerations are probabilistic and corresponding experiments,
Tversky and Shafir \cite{TS} and Shafir and Tversky \cite{ST}, are
statistical experiments.

\subsection{On applications of quantum formalism in psychology}

Already  Bohr pointed out \cite{BR}  to the possibility to apply
the mathematical formalism of quantum mechanics outside of
physics, in particular, in psychology. The {\it complementarity
principle} was considered as the starting point for application of
the quantum formalism outside of physics. Originally Bohr borrowed
this principle from psychology. Therefore he was sure that in turn
the formalism corresponding to  this principle could be applied to
psychology. We also mention a correspondence between Pauli and
Jung \cite{PJ}, \cite{PJ1} in the years 1932-1958.

Studies  of psychologist Wright \cite{Wright} on possibilities to
apply the quantum formalism to macroscopic (in particular,
cognitive systems) played an important role in understanding of
the probabilistic structure of quantum mechanics. The work of D.
Aerts and S. Aerts \cite{AR} stimulated applications of quantum
probability to psychology. It influenced essentially the author of
this paper. Quantum modelling in behavioral finances was performed
by Choustova \cite{Choustova0}, \cite{Choustova}  and Haven
\cite{Haven}. QL games approach to modelling of financial
processes was performed by  Piotrowski et al. \cite{PPP},
\cite{PP1}, see also Grib el al \cite{GR1} (and related works
\cite{GR4}, \cite{GR5}) for QL games for macroscopic players,
Danilov and Lambert-Mogiliansky \cite{DAN}, \cite{DAN1} applied
quantum logic type calculus of noncommutative actions to modelling
of decision making, in particular, in economics.

 We point out that the complementarity principle is a
general philosophical principle. In applications to quantum
physics it is quantatively exhibited through {\it interference
phenomenon} for discrete variables, see Dirac
\cite{Dirac}\footnote{Interference for continuous field-type
variables is well known in classical physics.} In purely
probabilistic terms interference can be represented as
interference of probabilities of alternatives. Detailed analysis
of this problem was performed in
\cite{Khrennikov/book:2004}--\cite{Khrennikov/Supp}. It was shown
that interference of probabilities can be represented as violation
of the {\it law of total probability} (also called the {\it law of
alternatives}) which is widely used  in classical statistics. This
effect was confirmed (at least preliminary) experimentally  by
Conte et al. \cite{Conte}.

Recently a similar viewpoint to the role of the law of total
probability  was presented by Busemeyer et al. \cite{Jerome1}, see
also \cite{F}, who described the well known disjunction effect
(violating Savage STP \cite{SV}) by using the quantum formalism,
see on this effect: Shafir and Tversky \cite{TS}, \cite{ST} and
also Rapoport \cite{RP},  Hofstader \cite{HOV}, \cite{HOV1} and
Groson \cite{G}.

\subsection{Law of total probability and its violations}

We recall this law in the simplest case of dichotomous random
variables, $a=\pm$ and $b=\pm$, see e.g.  wikipedia -- the article
``Law of total probability'': \begin{equation} \label{F} P(b=\pm)=
P(a=+) P(b=\pm\vert a=+) + P(a=-) P(b=\pm\vert a=-)
\end{equation}
Thus the probability $P(b=\pm)$ can be reconstructed on the basis
of conditional probabilities $P(b=\pm \vert a=\pm).$\footnote{
``The prior probability to obtain the result e.g. $b=+$ is equal
to the prior expected value of the posterior probability of $b=+$
under conditions $a=\pm.''$} This formula plays the fundamental
role in modern science. Its consequences are strongly incorporated
in modern scientific reasoning. It was a source of many scientific
successes, but at the same time its unbounded application induced
a number of paradoxes.\footnote{I think that the first paradox of
this type was disagreement between classical and quantum physics.}

In \cite{Khrennikov/book:2004}--\cite{Khrennikov/Supp} it was
pointed out that the quantum formalism induces a modification of
this formula. An additional term appears in the right hand side of
(\ref{F}), so called {\it interference term.} Violation of the law
of total probability  can be considered as an evidence that the
classical probabilistic description could not be applied (or if it
were applied, one could derive paradoxical conclusions). Our aim
is to show that QL probabilistic descriptions could be applied.
The terminology ``quantum-like'' and not simply ``quantum'' is
used to emphasize that violations of (\ref{F}) are not reduced to
those which can be described by the conventional quantum model.

Contexts which are nonclassical (in the sense of violation of
(\ref{F})), but at the same time cannot be described by the
conventional quantum formalism may appear outside quantum physics.
Nevertheless, the QL approach which was developed in
\cite{Khrennikov/book:2004}--\cite{Khrennikov/Supp} could be
applied even for such contexts (neither classical nor quantum).

\subsection{Mental contexts}

What are the sources of violation of the law of total probability?

\medskip

The most natural explanation can be provided in so called
contextual probabilistic framework
\cite{Khrennikov/book:2004}--\cite{Khrennikov/Supp}. The basic
notion of this approach is {\it context.} In quantum mechanics it
is a complex of experimental physical conditions.\footnote{The
notion of the context can be related to the notion of the {\it
preparation procedure} which is widely used in the quantum
measurement theory. Of course, preparation procedures -- devices
preparing  systems for subsequent measurements -- give a wide
class of contexts. However, the context is a more general concept.
For example, we can develop models operating with social,
political or historical contexts, e.g., socialism-context,
victorian-context. To give another example, one can consider the
context ``Leo Tolstoy" in literature.  The latter context can be
represented by various kinds of physical and mental systems -- by
books, readers, movies.} In the present paper it will be a complex
of mental conditions\footnote{The main terminological problem is
related to the notion of the {\it contextuality.} The use of the
term ``contextual'' is characterized by a huge diversity of
meanings, see  e.g. Svozil \cite{Svozil}, \cite{Svozil1} and
especially \cite{Svozil2}, \cite{Svozil3} for the notion of the
contextuality in quantum physics as well as Bernasconi and
Gustafson \cite{BG} for the notion of contextuality in cognitive
science and AI.}, see also \cite{Khrennikov/QLBrain}. In
particular, we shall consider contexts corresponding to {\it
Prisoner's Dilemma} (PD) as well as contexts for Tversky and
Shafir \cite{TS} gambling experiments. The crucial point is that
probabilities in the law of total probability correspond to
different contexts. A priori there is no reason to assume that all
those (essentially different contexts) could be ``peacefully
combined.'' Therefore in the contextual framework one could not
use {\it Boolean algebra} for contexts. We recall that Boolean
algebra is used in classical probability theory. It is important
to remark that in the latter conditioning is considered not with
respect to a context, but with respect to an event.

Roughly speaking violation of  the law of total probability  is
not surprising. It is surprising that we were able to find so many
situations (in particular, in classical statistical physics,
psychology and economics) in which it can be applied and that we
were lucky to proceed so far by using classical probability. The
latter can be explained if we consider this law as an {\it
approximative law.} If the additional term which should appear in
the general case in the right-hand side of (\ref{F}), the
``interference term'', see section 5, is relatively small, then
one could neglect by it and proceed by applying (\ref{F}) without
problem. In fact, the fundamental contribution of Tversky and
Shafir \cite{TS}, \cite{ST} is that they found statistical data
which {\it violates essentially} the law of total probability.

Our contextual approach does not contradict Bayesian approach
which nowadays is extremely popular in cognitive science and
psychology. We just say that Bayesian analysis is an {\it
approximative theory.} It has its domain of application. But (as
any mathematical model) it has its boundaries of application. From
our viewpoint the disjunction effect demonstrated that we have
approached these boundaries.

Thus the formula of total probability which is the basis of
Bayesian analysis is, in fact, not the precise equality (\ref{f}),
but it should be written as an approximative formula:
\begin{equation}
\label{FU} P(b=\pm)\approx P(a=+) P(b=\pm\vert a=+) + P(a=-)
P(b=\pm\vert a=-)
\end{equation}

\subsection{Numerical measure of mental interference}

In \cite{Khrennikov/book:2004}--\cite{Khrennikov/Supp} an {\it
interference coefficient}
 $\lambda$  was introduced. It gives a measure of incompatibility
of different contexts. It is important that this coefficient can
be found numerically on the
 basis of experimental statistical data. Moreover, by using this coefficient one can construct a quantum-like
(vector space) representation of contexts. Such a representation can be used e.g. in psychology or sociology.

Theoretical investigations of
\cite{Khrennikov/book:2004}--\cite{Khrennikov/Supp} demonstrated
that the situation is even
 more complicated than one might expect.
Besides the ordinary (well known)
trigonometric $\cos$-type interference (corresponding to the coefficient of interference
bounded by one), there exist incompatible contexts producing so called hyperbolic
$\cosh$-interference (corresponding to the coefficient of interference
larger than one). The latter type of probabilistic behavior could not be derived from
the conventional quantum mechanics. Such a hyperbolic interference has been never observed
for physical systems.

A cognitive experiment  which demonstrated that cognitive systems
(students) can behave in the QL way and produce nonzero
coefficients of interference was performed \cite{Conte}. It is
interesting that contexts (corresponding to Gestalt ambiguity
figures) used in this cognitive experiment produce the
coefficients of interference (providing a numerical measure of
incompatibility of these contexts) bounded by one.  Thus this
experiment on deviations of cognitive statistics from classical
statistics demonstrated the presence of usual trigonometric
interference -- as in classical and quantum wave mechanics.
Students behaved (with respect to recognition of Gestalt ambiguity
figures) in the same way as photons (with respect e.g. to choices
of slits in the two slit experiment -- diffraction of photons on
two slits).

\medskip

Can one hope to
find the hyperbolic interference in cognitive experiments?

\medskip

Intuitively there are no reasons to assume a priory that
incompatibility of contexts could not be so large that the
$\lambda$ would extend one. On the other hand, only the
trigonometric interference has been always produced in experiments
which have been done in classical and quantum physics. This as
well as the result of \cite{Conte} may induce opinion that the
hyperbolic interference is a kind of a theoretical artifact.

\subsection{Shafir-Tversky statistical effect}

Recently it was pointed out in Busemeyer et al. \cite{Jerome1} and
Franco \cite{F} that disjunction effects in cognitive sciences
could be explained on the basis of the quantum model. In the
present paper we shall continue their activity. We perform
QL-modeling of disjunction effects.

We apply the apparatus of contextual probability
\cite{Khrennikov/book:2004}--\cite{Khrennikov/Supp} to find
numerical characteristic -- the coefficient of interference (of
mental alternatives) for known experiments which demonstrated the
violation of the {\it sure thing principle} (Savage
\cite{SV}).\footnote{``If you prefer to compete knowing that your
opponent will compete {\it and} you prefer to to compete knowing
that your opponent will cooperate, then you should prefer to
compete even when you do not know yours opponent choice.''}

We shall use statistical data from Shafir and Tversky \cite{ST}
and Tversky and Shafir \cite{TS}. We find coefficients of mental
interference for these experiments. This will provide a
possibility to represent mental states of players (mental
contexts) by wave functions -- in the abstract approach by
normalized vectors of Hilbert space.
 We recall that in \cite{Khrennikov/book:2004}--\cite{Khrennikov/Supp} an algorithm for such a representation was presented,
{\it Quantum-Like Representation Algorithm, QLRA.}

We found not only that probabilistic behaviors are nonclassical in
both experiments (this was already shown in  Busemeyer et al.
\cite{Jerome1}), but that they differ essentially. We found that
Tversky and Shafir \cite{TS} experiment produces the conventional
trigonometric interference and consequently players behave under
game-contexts in the same way as photons behave under contexts of
the two slit experiment. Surprisingly Shafir and Tversky \cite{ST}
experiment does not (!) produce the conventional trigonometric
interference. It produces one interference coefficient which is
larger than 1 -- hyperbolic interference, and another which is
less than 1 -- trigonometric interference.

\medskip

Thus Shafir and Tversky \cite{ST} experiment produces the
hyper-trigonometric interference!

\medskip

This is the first experimental evidence of hyperbolic
interference! And it was found not in physics, but in cognitive
science.

\subsection{Quantum-like thinking}

As we pointed out, in
\cite{Khrennikov/book:2004}--\cite{Khrennikov/Supp}  an algorithm,
QLRA, for mapping probabilistic data into linear space of
probability amplitudes was proposed. It represents contextual
probabilities by wave functions (or normalized vectors of Hilbert
space). We speculate, see also \cite{Khrennikov/QLBrain},  that
cognitive systems might develop (in the process of mental
evolution) the ability to apply QLRA and to create
QL-representations of mental contexts. Thus, instead of operating
with probabilities and analyzing (even unconsciously)
probabilities of various alternatives, the brain works directly
with mental wave functions (probabilistic amplitudes).

Such a QL-processing of information has the following advantages:

\medskip

a). This is consistent processing of {\it incomplete information.}
The crucial point is that it is {\it consistent information cut off.} Therefore
such a processing does not induce ``information chaos'', especially under the assumption that
all cognitive systems use the same QL-representation.

\medskip

b). This is linear (vector space) processing of information. From the purely mathematical viewpoint
one can consider this procedure as {\it linearization of probabilistic representation of mental contexts.}
In particular, the mental wave function evolves linearly. Such an evolution is described
by mental Schr\"odinger's equation.\footnote{Thus we guess that the brain was able to linearize
the mental world via the QL-representation.}

\medskip

We speculate that the biological evolution induced  the QL-representation
of information  long before discovery of  quantum mechanics by Planck, Einstein, Bohr, Heisenberg,
Schr\"odinger, Dirac, von Neumann.

We also emphasize that our hypothesis on QL-processing of mental
information has nothing to do with so called {\it quantum
reductionist theories,} e.g. \cite{Hameroff1},  \cite{Hameroff2},
\cite{Penrose1}, \cite{Penrose2}. By the latter processing of
information by cognitive systems have some quantum features,
because the brain (as any physical system) is composed of quantum
particles. Yes, the brain is composed of e.g. electrons, protons
and photons, but this has nothing to do with QL-representation of
mental contexts which is performed on the macro level.\footnote{We
remark that neuronal models of QL-representation of mental
information have not yet been developed. But we expect that our QL
cognitive modeling may stimulate neurophysiological studies.}

\subsection{Quantum-like decision making}

If our hypothesis on QL-processing of information by cognitive systems is correct, then we should
consider the QL-process of decision making. We recall that decision making is the cognitive process
leading to the selection of a course of action among variations. Every decision
making process produces a final choice.

By our model a cognitive system represents a mental context, say
$C,$ underlying decision making by a mental wave function,
probabilistic (complex or even hyperbolic) amplitude $\psi_C.$ This
mental wave function evolves linearly in the Hilbert state space:
$\psi_C(t).$ `Decision making operation'' is represented by an
observable, say $b,$ taking values corresponding to different
choices of action. Its value corresponding to the choice between
alternatives  is generated (by a classical random generator)  with
the probability given by the Born's rule for the mental wave function $\psi_C(T),$  where
$T$ is the instance of time corresponding to decision making.

On the basis of such a QL-representation this cognitive system
selects a course of action among variations {\it purely
automatically} (i.e., without applying the rule of reason based on
the conventional Boolean logic) on the basis of a random generator
reproducing the probability distribution of the QL-observable $b$
for the wave function $\psi_C(T).$  This probability distribution
is given by Born's rule.\footnote{We remark that in
\cite{Khrennikov/book:2004}--\cite{Khrennikov/Supp} Born's rule
was generalized even to QL models which differ from the
conventional quantum model.}

\medskip

To get the probability that an observable $b$  takes a fixed value,
the brain should find the scalar product of the wave function $\psi_C(T)$ and the eigenvector
corresponding this value. Finally, the absolute value of the
result of this procedure should be squared.

\medskip

Thus we assume that (at least some) cognitive systems have following QL-abilities:

\medskip

a). To apply QLRA and to create the QL-representation of mental
contexts: a context $C$ is mapped into its wave function $\psi_C;$

b). To generate dynamics of the mental wave function described by Schr\"odinger equation;

c). To represent ``decision making observables'' by linear
operators\footnote{Since decision's spectrum consists of discrete
alternatives, it is enough to operate in finite dimensional linear
spaces, i.e., with matrices. In quantum mechanics observables are
represented by self-adjoint operators, i.e., by symmetric
matrices. However, we speak not only about conventional quantum
representation of cognitive entities, but about QL-representation
which is based on the contextual approach. As was found in
\cite{Khrennikov/book:2004}--\cite{Khrennikov/Supp}, contextual
probabilistic setups could  violate not only the classical
probabilistic laws, for example, the law of total probability, but
even the conventional quantum laws. For example, it might happen
that a mental observable could not be represented by a symmetric
matrix.};

d). To apply Born's rule and to create random generators for probability distributions
based on this rule.

\medskip

As was already pointed out in footnote 8, the QL-representation is essentially more general than
the conventional quantum representation. For example, some mental contexts might be represented not by complex
probability amplitudes, but by hyperbolic (or even mixed hyper-trigonometric) amplitudes.

\subsection{Quantum-like superposition of choices}

We remark that QL decision making also includes the QL-dynamics of the mental state
$\psi_C.$  Of course, in the same way as in the
conventional quantum mechanics by making a concrete choice among alternatives a cognitive system
disturbs the QL-evolution which is described (at least approximately) by Schr\"odinger's equation.

One could say that ``collapse of the mental wave function'' occurs
at the instant of time $t=T.$ In opposite to the conventional
Copenhagen interpretation, we do not take collapse too seriously.
In our model the $\psi_C$-function is simply a special linear
space representation of probabilistic data about the context $C.$
In the process of decision making the (self) measurement of a
decision maker $b$ is realized in purely classical way. It is
assumed existence (in the brain) of a random generator which
produces possible values of $b$ with probabilities given by Born's
rule. Let e.g. $b$ take two values. These are two alternative
decisions: +1, yes, or -1, no. Then the mental wave function and
the decision maker determine two probabilities, $p_+$ and $p_-.$
The values $b=+ 1$ and $b=- 1$ appear randomly with these
probabilities.

Suppose that a cognitive system should make the $b$-decision. This
system runs the above random generator. It takes the value $b=+
1.$  At this moment the Schr\"odinger evolution is
stopped. It starts again with a new initial mental wave function
which is equal to the eigenvector corresponding to the value $b=+
1.$ In accordance with quantum terminology we can say that during
the period $0 \leq t \leq T$ the brain's mental state was in the
superposition of two states $b=+1$ and $b=-1.$

In section 1.9 we shall consider more complicated process: a new
context can be formed and represented by its own mental wave
function. Evolution may start with it and not with the eigenvector
corresponding to the previous decision.

In general  a mental context $C$ can be created not specially for
making the $b$-decision.  Decision tasks can come later.
Suppose that the brain has a collection of decision
makers (self-observables) $a,b,...$\footnote{It may be better to
consider ``activated decision makers''. The total number of possible
decision makers can be essentially larger. However, majority of
them are in the ``sleeping state.''} The mental wave function
$\psi_C(t)$ can be considered (by the conventional quantum
terminology) as being in superposition of all possible values for
any observable. If the cognitive system should make the
$b$-decision, then the $b$-superposition is reduced to a single
value, e.g. $b=+1.$ Suppose that operators (matrices) representing
observables $a$ and $b$ do not commute. Then the eigenvector of
$b$ for the value $b=+1$ need not be at the same time an eigenvector for
$a.$ Hence, after taking the decision $b=+1$ the brain's state is
still in superposition of all possible values for the $a.$

Although we use the same terminology as in quantum mechanics,
states' superposition, its interpretation is totally different
from the conventional one. Therefore we prefer to speak about QL
superposition of mental states and not quantum superposition. The
first is described in purely classical terms (even Schr\"odinger's
dynamics can be easily simulated by classical neural network).
Therefore it can be exhibited by {\it macroscopic systems.} The
original quantum superposition is ``real superposition'' of e.g.
two energy levels. It is not clear how it might be realized for
macroscopic systems. The model of the brain operating with quantum
superpositions of minds is very old. It was proposed by quantum
logician Vladimir Orlov \cite{OR} (in fact, a few years earlier,
but it took time to transfer the manuscript from a concentration
camp for decedents). Similar model was considered by Stuart
Hameroff \cite{Hameroff1},  \cite{Hameroff2} and Roger Penrose
\cite{Penrose1}, \cite{Penrose2}. But they understood well the
problem, see e.g. Roger Penrose  \cite{Penrose2}: {\small ``It is
hard to see how one could usefully consider a quantum
superposition consisting of one neuron {\it firing,} and
simultaneously {\it nonfiring.}''}

\subsection{Parallelism in creation and processing of mental
function}

It is clear that the brain cannot operate for a long time starting
with some context $C.$ A series of Schr\"odinger's evolutions and
``state updating'' after decision making can be stopped as a
consequence of creation of a new mental context $C^\prime$ induced
by new external and internal signals. This context is represented
by its own mental wave function $\psi_{C^\prime}$ which evolves
linearly in the Hilbert state space. The process of decision
making and state updating is repeated starting with
$\psi_{C^\prime}.$

If the brain's evolution was done properly from the
point of view of the information processing architecture, then it is
natural to assume that  creation of a new context and
its QL representation can go in parallel to processing, decision making and state
updating based on the previous context $C.$

We consider two domains of the brain, classical  and QL. In
principle, each domain can be distributed through the brain (for example, if the
neural basis is given by the frequency domain representation).

In the classical domain a probabilistic image of a mental context
$C$ is created.\footnote{As was pointed out in
\cite{Khrennikov/book:2004}, \cite{KHN}, probabilities may be
generated by counting frequencies of neural firings. However, such
a model is just a possible candidate for the neuronal basis. In
any event extended neurophysiological investigations should be
performed to find mechanism of neural creation of the QL
representation and dynamics as well as self-measurements in the
process of decision making.}  Then these contextual probabilities
are represented by the mental wave function.

This mental wave function is processed in the QL domain:
Schr\"odinger's evolution, measurement, updating, and so on.

The classical domain does not ``sleep'' meanwhile. It works with a
new context, say $C^\prime.$  Its amplitude representation will be
transferred to the QL domain later.

There should be a king of control center coupling consistently
functioning of these two domains. In particular, it should control
consistency of time scales for state preparation and decision
making. On the one hand, the brain saves a lot of computational
resources by working only in the QL domain. Here dynamics is
linear -- in opposite to essentially nonlinear dynamics in the
classical domain of the brain. However, new signals change mental
context and it should be updated (in the classical domain).

\subsection{Quantum-like rationality}

If one defines rational behavior on the basis of the law of total
probability, then QL-behaviors would be really irrational, see
section 2 on rational behavior, PD and so on. However, the only
reason for such an interpretation is common application of the law
of total probability in modern statistics. Under the assumption
that cognitive systems make decisions via the  QL decision making
procedure, violation of ``Boolean rationality'' does not look
surprising. One must be essentially more surprised that modern
science (including economy and finances) was able to proceed so
far on the basis of assumptions based on classical ``Boolean
rationality.''

Therefore one should consider deviations from ``Boolean rationality'' not as evidences of irrational
behavior, but as evidences that cognitive systems are QL-rational.

We point out to another source of QL-rationality. Besides advantages of QL-processing of incomplete
information, see section 1.6, we mention presence of  social pressure
to proceed in the QL-way.
If society consists of QL-thinking cognitive systems, then any individual should use
the QL-reasoning to proceed consistently with respect to other members of such a QL-society.
An individual who tries to use  essentially more detailed description of mental contexts
and who tries to build classical-like complete representation of  contexts could make
decisions which are in fact ``more rational'' (from the point of view of complete
information processing). However, such an individual might be rejected
by the QL-society.

\subsection{Quantum-like ethics}

We remark that ``nonconsequential reasoning'' was studied a lot in
cognitive psychology, e.g. Rapoport \cite{RP},  Hofstader
\cite{HOV}, \cite{HOV1}, Tversky and Shafir \cite{TS}, \cite{ST}.
However, from the QL point of view such a reasoning is not
nonconsequential at all. It is consequential, but consequences are
taken into account in the QL-representation. For example,
preference of cooperative, ethical decisions in PD  is
consequential, but from the viewpoint of QL probability. Hence,
human ethics is fact a consequence of the QL-representation of
mental contexts. If we were involved in purely classical
probabilistic reasoning (based on classical Baeysian analysis), we
would not be able to demonstrate such a ``nonconsequential
behavior'' as in PD. We would behave as ``cognitive automata'' (as
creations of AI).  The essence of human behavior is the presence
of the QL-representation of probabilistic reality. Cooperation may
arise simply because the mental wave function produces (via Born's
rule) larger probabilities for cooperative actions.

In the absence of decision making the mental wave function evolves
according to Schr\"odinger's equation. The generator of evolution
is represented by a special QL observable -- ``mental
Hamiltonian'', describing a mental analogue of energy, see
\cite{KhrennikovAN}, \cite{KhrennikovF} for details.

We guess that human beings have  mental Hamiltonians such that they  produce ``ethical wave functions'', $\psi_C(T),$
starting with a large variety of $\psi_C.$ Creation of such ``ethic mental Hamiltonian'' is a consequence of influence of
social environment already in childhood. We could not exclude that some terms of ``ethic mental Hamiltonian''
are encoded in genom.

\section{Rational behavior, Prisoner's Dilemma}

In game theory, PD is a type of non-zero-sum game in which two
players can cooperate with or defect (i.e. betray) the other
player. In this game, as in all game theory, the only concern of
each individual player (prisoner) is maximizing his/her own
payoff, without any concern for the other player's payoff. In the
classic form of this game, cooperating is strictly dominated by
defecting, so that the only possible equilibrium for the game is
for all players to defect. In simpler terms, no matter what the
other player does, one player will always gain a greater payoff by
playing defect. Since in any situation playing defect is more
beneficial than cooperating, all rational players will play
defect.

The classical PD is as follows: Two suspects, $A$ and $B,$ are
arrested by the police. The police have insufficient evidence for
a conviction, and, having separated both prisoners, visit each of
them to offer the same deal: if one testifies for the prosecution
against the other and the other remains silent,
 the betrayer goes free and the silent accomplice receives the full 10-year sentence.
If both stay silent, both prisoners are sentenced to only six
months in jail for a minor charge. If each betrays the other, each
receives a two-year sentence. Each prisoner must make the choice
of whether to betray the other or to remain silent. However,
neither prisoner knows for sure what choice the other prisoner
will make. So this dilemma poses the question: How should the
prisoners act? The dilemma arises when one assumes that both
prisoners only care about minimizing their own jail terms. Each
prisoner has two options: to cooperate with his accomplice and
stay quiet, or to defect from their implied pact and betray his
accomplice in return for a lighter sentence. The outcome of each
choice depends on the choice of the accomplice, but each prisoner
must choose without knowing what his accomplice has chosen to do.
In deciding what to do in strategic situations, it is normally
important to predict what others will do. {\it This is not the
case here.} If you knew the other prisoner would stay silent, your
best move is to betray as you then walk free instead of receiving
the minor sentence. If you knew the other prisoner would betray,
your best move is still to betray, as you receive a lesser
sentence than by silence. Betraying is a dominant strategy. The
other prisoner reasons similarly, and therefore also chooses to
betray. Yet by both defecting they get a lower payoff than they
would get by staying silent. So rational, self-interested play
results in each prisoner being worse off than if they had stayed
silent, see e.g. wikipedia -- ``Prisoner's dilemma.''

This is the {\it principle of rational behavior} which is basic
for rational choice theory which is the dominant theoretical
paradigm in microeconomics. It is also central to modern political
science and is used by scholars in other disciplines such as
sociology. However, Shafir and Tversky \cite{ST} found that
players frequently behave irrationally.

\section{Contextual analysis of Prisoner's Dilemma}

Each contextual model is based on a collection of contexts and a
collection of observables. Such observables can be
measured\footnote{By measurements we understand even
self-measurements which are performed by e.g. the brain.} for each
of  contexts under consideration, see \cite{Khrennikov/Supp} for
the general formalism. The following mental contexts are involved
in PD:

\medskip

Context $C$ representing the situation such that a player has no
idea about planned action of another player.

\medskip

Context $C_{+}^A$ representing the situation such that the
$B$-player supposes that $A$ will cooperate and context $C_{-}^A$
-- $A$ will compete. We can also consider similar contexts
$C_{\pm}^B.$

\medskip

We define dichotomous observables $a$ and $b$ corresponding to
{\it actions} of players $A$ and $B:$ $a=+$ if $A$ chooses to
cooperate and $a=-$  if $A$ chooses to compete, $b$ is defined in
the same way.

A priory the law of total probability might be violated for PD,
since the $B$-player is not able to combine contexts. If those
contexts were represented by subsets of a so called space of
``elementary events'' as it is done in classical probability
theory (based on Kolmogorov (1933) measure-theoretic axiomatics),
the $B$-player would be able to consider the conjunction of the
contexts $C$ and  e.g. $C_{+}^A$ and to operate in the context $C
\wedge C_{+}^A$ (which would be represented by the set $C \cap
C_{+}^A).$ But the very situation of PD is such that one could not
expect that contexts $C$ and  $C_{\pm}^A$ might be peacefully
combined. If the $B$-player obtains information about the planned
action of the $A$-player (or even if he just decides that $A$ will
play in the definite way, e.g. the context $C_{+}^A$ will be
realized), then the context $C$ is simply destroyed. It could not
be combined with $C_{+}^A.$

We can introduce the following contextual probabilities:

\medskip

$P(b=\pm \vert C)$ -- probabilities for actions of $B$ under the
complex of mental conditions $C.$

\medskip

$P_{\pm,+}\equiv P(b=\pm \vert C_+^A)$ and $P_{\pm,-}\equiv
P(b=\pm \vert C_-^A)$ -- probabilities for actions of $B$ under
the complexes of mental conditions $C_+^A$ and $C_-^A,$
respectively.

\medskip

$P(a=\pm \vert C)$ -- priory probabilities which $B$ assigns  for
actions of $A$ under the complex of mental conditions $C.$

\medskip

As we pointed out, there are no priory reasons for the equality
(\ref{F}) to hold. And experimental results of Shafir and Tversky
\cite{ST} demonstrated that this equality could be really
violated, see Busemeyer et al. \cite{Jerome1}.

By Shafir and Tversky \cite{ST} for PD experiment we have:

\medskip

$P(b=- \vert C)=0.63$ and hence $P(b=+ \vert C)=0.37;$

\medskip

$P_{-,-}=0.97, \; P_{+,-}=0.03;\; \;P_{-,+}=0.84, \; P_{+,+}=0.16.$

\medskip

As always in probability theory it is convenient to introduce the matrix of
transition probabilities
$$
P = \left( \begin{array}{ll}
0.16 & 0.84\\
0.03& 0.97\\
\end{array}
\right ).
$$
We point out that this matrix is {\it stochastic.} It is
a square matrix each of whose rows consists of nonnegative real numbers,
with each row summing to 1. This is the common property of all matrices of transition
probabilities.

We now recall the definition of a {\it doubly stochastic matrix:}
in a doubly stochastic matrix all entries are nonnegative and all
rows and all columns sum to 1. It is clear that the {\it matrix
obtained by Shafir and Tversky is not doubly stochastic.}

In the simplified framework the prisoner $B$ considers (typically
unconsciously) priory probabilities  $p= P(a=+ \vert C)$ and $1-p=
P(a=- \vert C)$ which $B$ assigns for actions of $A$ under the
complex of mental conditions $C.$ These probabilities are
parameters of the model. In the simplest case $B$ assigns some
fixed value $p$ to  $A$-cooperation. The mental wave function
depends on $p.$

However, in reality the situation is essentially more complicated.
The $B$ is not able to determine precisely $p.$ He considers a
spectrum of possible $p$ which might be assigned to
$A$-cooperation. Therefore, instead of a pure QL-state (mental
wave function), the $B$-brain creates a {\it statistical mixture
of mental wave functions} corresponding to some range of
parameters $p$ which could be assigned to $A$-cooperation. In this
statistical mixture different wave functions are mixed with some
weights. Instead of the wave function, $B$ creates a von Neumann
density matrix which describes $B$'s state of mind. We emphasize
that the latter operation of statistical mixing is purely
classical. The crucial step is creation of the QL-representation
for fixed value of the parameter $p.$

\section{Contextual analysis for Tversky and Shafir  gambling experiment}

Tversky and Shafir \cite{TS} proposed to test disjunction effect
for the following gambling experiment. In this experiment, you are
presented with two possible plays of a gamble that is equally
likely to win 200 USD or lose 100USD. You are instructed that the
first play has completed, and now you are faced with the
possibility of another play.

Here a gambling device, e.g., roulette,  plays the role of $A;$
$B$ is a real player, his actions are  $b=+,$ to play the second
game, $b=-,$  not. Here the context $C$ correspond to the
situation such that the result of the first game is unknown for
$B;$ the contexts $C_{\pm}^A$ correspond to the situations such
that the results  $a=\pm$ of the first play in the gamble are
known.

From Tversky and Shafir \cite{TS} we have:

\medskip

$P(b=+ \vert C)=0.36$ and hence $P(b=- \vert C)=0.64;$

\medskip

$P_{+,-}=0.59, \; P_{-,-}=0.41;\; \; P_{+.+}=0.69, \; P_{-.+}=0.31.$

\medskip

We get the following matrix of transition probabilities:
$$
P = \left( \begin{array}{ll}
0.69 & 0.31\\
0.59& 0.41\\
\end{array}
\right ).
$$
This matrix of transition probabilities is neither (cf.
Shafir-Tversky \cite{ST} experiment) doubly stochastic.

In this experiment (in contrast to Shafir-Tversky \cite{ST})
probabilities $P(a=\pm \vert C)$ are not subject of a priory
consideration. They are fixed from the very beginning as 1/2.

\section{Coefficient of interference (incompatibility)}

Violation of the law of total probability implies that the left-hand and right-hand sides of
(\ref{F}) do not coincides. Therefore it is natural to consider the difference between them
as a measure of incompatibility between contexts $C$ and $C_A^{\pm}.$
We denote it by the symbol $\delta_{\pm}.$
It is the measure of impossibility
to combine these contexts in a single space of elementary events. In PD $C$ can be called
uncertainty context -- $B$ has no information about planned actions of $A.$ This context is incompatible
with the contexts $C_A^{\pm}$ corresponding to definite actions of $A.$ We propose
to measure this incompatibility numerically  by using $\delta.$ This number
can be found if one have all probabilities involved in the law of total probability.

The next important question is the choice of normalization of
$\delta.$ Here we proceed in the following way, see
\cite{Khrennikov/book:2004}. We are lucky that quantum mechanics
has been already discovered. Its formalism implies \cite{Dirac}
that for quantum systems (e.g. photons) this coefficient of
incompatibility has the form $2 \cos \theta$ (where the angle
$\theta$ is called phase) multiplied by the normalization factor
which is equal to square root of the product $\Pi$ of all
probabilities in the right-hand side of (\ref{F}). Thus
$$
\delta= 2 \cos \theta \; \sqrt{\Pi}.
$$
We proposed to use the same normalization in the general case of
any collection of contextual probabilities.\footnote{We remark
that in general we could expect neither classical nor conventional
quantum probabilistic behaviors.} Thus we introduce the normalized
coefficient of incompatibility of mental contexts:
$$
\lambda= \frac{\delta}{2\; \sqrt{\Pi}}.
$$
As was mentioned, in the conventional quantum mechanics it is always bounded by one.
Hence, it can be written
as $\lambda =\cos \theta,$ where $\theta= \arccos \lambda.$

However, as was found in \cite{Khrennikov/Supp}, it could as well
be larger than one. In such a case it can be written as $\lambda =
\pm \cosh \theta,$ where $\theta= \rm{arccosh}\;  \vert \lambda
\vert .$

Since in the conventional quantum mechanics the term $\delta= 2 \cos \theta \; \sqrt{\Pi}$ describes
interference,  we can call $\delta$ the interference term even in the general contextual framework.
 The same terminology
we use for the normalized coefficient  $\lambda:$
the coefficient of interference.
 It can be considered as a measure of
{\it ``interference of mental contexts.''}

\section{Coefficients of interference for disjunction experiments}

Since in the Tversky and Shafir \cite{TS}  gambling experiment the
$A$-probabilities are fixed, it is easier for investigation.
Simple arithmetic calculations give $\delta_+= -0.28,$ and hence
$\lambda_+= -0.44.$ Thus the probabilistic phase $\theta_+= 2.03.$
We recall  \cite{Khrennikov/book:2004} that $\delta_+ +
\delta_-=0$ (in the general case). Thus $\delta_-= 0.28,$ and
hence $\lambda_-= 0.79.$ Thus the probabilistic phase $\theta_-=
0.66.$

In the case of Shafir and Tversky \cite{ST} PD-experiment the
$B$-player assigns probabilities of the $A$-actions, $p$ and $1-p$
(in the simplest case). Thus coefficients of interference depend
on $p.$ We start with $\delta_-= -(0.21 +0.13 p)$ and $\lambda_-=
-(0.12 +0.07 p)/\sqrt{p(1-p)}.$ For example, if $B$ would assume
that $A$ will act randomly with probabilities $p=1-p=1/2,$ then
the interference between contexts is given by $\lambda_- = -0.31$
and hence the phase $\theta= 1.89.$ We now find $\delta_+= (0.21
+0.13 p)$ and $\lambda_+= -(1.52 + 0.94 p)/\sqrt{p(1-p)}.$ For
example, if $B$ would assume that $A$ will act randomly with
probabilities $p=1-p=1/2,$ then the interference between contexts
is given by $\lambda_+=3.98.$ Thus interference is very high. It
exceeds the possible range of the conventional trigonometric
interference. This is {\it the case of hyperbolic interference!}
Here the hyperbolic phase $\theta_+=  \rm{arccosh}\; (3.98)=
2.06.$

\section{Quantum-like representation algorithm - QLRA}

This algorithm will produce a probability amplitude from
contextual probabilities. We shall consider separately two cases:

\subsection{Trigonometric mental interference}

The coefficients of interference are bounded by one.

In this case we can represent $\lambda_{\pm}$ in the form
$\lambda_{\pm} = 2 \cos \theta_{\pm} \sqrt{\Pi}.$ Hence we obtain the following modification of the
law of total probability:
\begin{equation}
\label{F1}
P(b=\pm)= P(a=+) P_{\pm,+} + P(a=-) P_{\pm,-}+ 2 \cos \theta_{\pm} \sqrt{\Pi},
\end{equation}
where $\Pi_{\pm}= P(a=+\vert C) P(a=-\vert C)P_{\pm,+}P_{\pm,-}.$
In a special case -- for a doubly stochastic matrix of transition probabilities --
this law can be derived in the conventional quantum formalism.

We now recall elementary formula from algebra of complex numbers:
$$
k=k_1+k_2+2\sqrt{k_1k_2}\cos \theta=\vert \sqrt{k_1}+e^{i \theta}\sqrt{k_2}|^2,
$$
for real numbers $k_1, k_2 > 0, \theta\in [0,2 \pi].$
Thus
$$
k= \vert \psi |^2, \; \mbox{where}\; \psi=\sqrt{k_1}+e^{i \theta}\sqrt{k_2}.
$$
Let us compare this formula and the interference law of total probability
(\ref{F1}). We set $k=P(b=\pm), k_1= P(a=+) P_{\pm,+}, k_2= P(a=-) P_{\pm,-}.$
We introduce the complex probability amplitudes:
$$
\psi(\pm)= \sqrt{P(a=+) P_{\pm,+}} + e^{i \theta_{\pm}} \sqrt{P(a=-) P_{\pm,-}}.
$$
We call its mental wave function (it is defined on the set $\{+, -\}$ and takes
complex values) representing the context $C$ via observables $a$ and $b.$

The crucial point is that Born's rule takes place:
$$
P(b=\pm)= \vert \psi(\pm) |^2.
$$
We speculate that the brain can apply such an algorithm to
probabilistic data about contexts and construct the complex
probability amplitude, the mental wave function. Then it operates
only with such amplitudes and not with original probabilities.

\subsection{Hyperbolic mental interference}

The coefficients of interference are larger than one.

Here mathematics is more complicated. One should use so called
hyperbolic numbers, instead of complex numbers. We would not like
to go in mathematical details. We just mention that one should
change everywhere the imaginary unit $i$ (such that $i^2=-1)$ to
hyper-imaginary unit $j: j^2=+1$  and usual trigonometric
functions $\cos \theta$ and $\sin \theta$ to their hyperbolic
analogues $\cosh \theta$ and $\sinh \theta,$ see e.g.
\cite{Khrennikov/Supp} for details. Here the probabilistic image
of incompatible mental contexts is given by the hyperbolic
probabilistic amplitude:
$$
\psi(\pm)= \sqrt{P(a=+) P_{\pm,+}} \pm e^{j \theta_{\pm}} \sqrt{P(a=-) P_{\pm,-}}.
$$

Finally, we remark that some cognitive systems may exhibit (for some mental contexts)
hyper-trigonometric interference: one coefficient, e.g., $\lambda_+$ is bounded by one
and another is larger than one.

\subsection{Quantum-like representation for Tversky-Shafir  gambling experiment}

This experiment has simpler QL-representation. Both coefficients of interference are bounded
by one. Thus we can represent incompatible contexts by the complex probability amplitude:
$$
\psi(+) \approx 0.59+ e^{2.03 i} \;  0.54;\; \; \; \psi(-)\approx 0.39+ e^{0.79 i} \; 0.45.
$$

\subsection{Quantum-like representation for Shafir-Tversky  PD experiment}

Here the $B$-player creates QL-representation by assigning the probabilities
$p$ and $1-p$ to possible actions of $A.$ The wave function depends on $p.$
For example, suppose that $B$ assigned to the $A$-actions equal probabilities.
Then the $B$-brain would represent the PD game by the following hyper-trigonometric
amplitude:
$$
\psi(+)\approx 0.28 + e^{2.06 j} \; 0.12;\; \; \; \; \psi(-)\approx  0.65 + e^{1.89i}\; 0.7
$$

\section{Non-doubly stochasticity of matrices
of transition probabilities in cognitive science}

We have seen that matrices of transition probabilities  which are
based on experimental data of Tversky-Shafir  Game and
Shafir-Tversky PD experiments are not doubly stochastic. The same
is valid for the matrix obtained in the Bari-experiment
\cite{Conte}. On the other hand, matrices of transition
probabilities that should be generated by the conventional quantum
mechanics in the two dimensional Hilbert space are always doubly
stochastic, see \cite{VN}.

We can present two possible explanations of this ``non-doubly
stochasticity paradox'':

a). Statistics of these experiments are neither classical  nor
quantum (i.e., neither the Kolmogorov measure-theoretic model nor
the conventional quantum model with self-adjoint operators could
describe this statistics).

b). Observables corresponding to real and possible actions are not
complete. From the viewpoint of quantum mechanics this means that
they should be represented not in the two dimensional (mental
qubit) Hilbert space, but in Hilbert space of a higher dimension.
\footnote{This latter possibility was pointed to me by Jerome
Busemeyer and Ariane Lambert-Mogiliansky during the recent
workshop ``Can quantum formalism be applied in psychology and
economy?'' (Int. Center for Math. Modeling in Physics and
Cognitive Sciences, University of V\"axj\"o, Sweden; 17-18
September, 2007).}

Personally I would choose the a)-explanation (and not simply
because it was my own). It seems that actions of $A$ and $B$ in
the PD do not have a finer QL-representation which would be
natural with respect to the QL-machinery of decision making.

Of course, there are many brain-variables which are involved in
the PD decision making. However, the essence of creation of a QL
representation is  selection of the most essential variables.
Other variables should not be included in the chosen (for a
concrete problem) QL representation.

Nevertheless, we could not ignore completely the incompleteness
conjecture of Busemeyer and Lambert-Mogiliansky. We would
immediately meet a really terrible problem: ``How can we find the
real dimension of the quantum (or QL) state space?'' So, if this
dimension is not determined by values of complementary observables
$a$ and $b,$ then we should be able to find an answer to the
question: ``Which are those additional mental observables which
could complete the model?'' One should find complete families of
observables $u_1^a,..., u_m^a$ and $u_1^b,..., u_m^b.$ compatible
with $a$ and $b,$ respectively.

We remark that in the case of the hyperbolic interference we would
not be able to solve the ``non-doubly stochasticity paradox'' even
by going to higher dimensions.

\medskip

My conjecture (similar ideas also were presented by Luigi Accardi
and Dierk Aerts, at least in conversations with me and our  Email
exchange) is that the laws of classical probability theory can be
violated in cognitive sciences, psychology, social sciences and
economy. However, nonclassical statistical data is not covered
completely by the conventional quantum model.

My personal explanation is based on the evidence
\cite{Khrennikov/book:2004} that violation of the formula of total
probability does not mean that we should obtain precisely the
formula of total probability with the interference term which is
derived in the conventional quantum formalism.

Nevertheless, the conventional quantum formalism can be used as
the simplest nonclassical model for mental and social modelling.

\medskip

{\bf Conclusion.} {\it By using violation of the law of total
probability as the starting point we created the QL-representation
of mental contexts. As was pointed out by Busemeyer et al.,
violation of the law of total probability can be used to explain
disjunction effect. Therefore the QL-representation can be applied
for description of this effect. The essence of our approach is the
possibility to introduce a numerical measure of disjunction, so
called interference coefficient.  In particular, we found
interference coefficients for
 statistical data from Shafir--Tversky  and Tversky--Shafir
experiments coupled to Prisoner's Dilemma. We also represent
contexts of these experiments by QL probability amplitudes,
``mental wave functions.'' We found that, besides the conventional
trigonometric interference (Tversky--Shafir \cite{TS}) in
cognitive science can be exhibited so called hyperbolic
interference - Shafir and Tversky \cite{ST}. Thus the
probabilistic structure of cognitive science is not simply
nonclassical (cf. \cite{Conte}, \cite{Jerome1}), but it is even
essentially richer than the probabilistic structure of quantum
mechanics.}

\end{document}